\documentstyle[aps,epsfig]{revtex}

\newcommand{\zaa}{Astron.~Astrophys.}

\newcommand{\zapj}{Astrophys.~J.}
\newcommand{\zapjl}{Astrophys.~J.~Lett.}
\newcommand{\zapjs}{Astrophys.~J.~S.}
\newcommand{\znas}{New~Astronomy}
\newcommand{\znat}{Nature}
\newcommand{\zepj}{European Physical Journal}
\newcommand{\znp}{Nucl.~Phys.}

\newcommand{\zpr}{Phys.~Rev.}
\newcommand{\zprl}{Phys.~Rev.~Lett.}

\newcommand{\zadndt}{At. Data Nucl. Data Tables}
\newcommand{\zmnras}{Mon. Not. R. Astron. Soc.}

\newcommand{\gap}{\mathrel{ \rlap{\raise.5ex\hbox{$>$}}
                    {\lower.5ex\hbox{$\sim$}}  } }
\newcommand{\lap}{\mathrel{ \rlap{\raise.5ex\hbox{$<$}}
	            {\lower.5ex\hbox{$\sim$}}  } }
\newcommand{\ob}{$\Omega_b$}
\newcommand{\obh}{$\Omega_bh^2$}
\newcommand{\lyma}{Lyman--$\alpha$}
\newcommand{\deu}{$D$}
\newcommand{\tro}{$^3He$}
\newcommand{\qua}{$^4He$}
\newcommand{\six}{$^{6}Li$}
\newcommand{\sep}{$^{7}Li$}

\begin{document}
\twocolumn

\title{Constraints on \ob\ from the nucleosynthesis of \sep\
in the standard Big Bang}

\author{Alain Coc}

\address{
Centre de Spectrom\'etrie Nucl\'eaire et de Spectrom\'etrie
de Masse, IN2P3-CNRS and Universit\'e Paris Sud, B\^atiment 104,\\
91405 Orsay Campus, France
}

\author{Elisabeth Vangioni-Flam}
\address{
Institut d'Astrophysique de Paris, 98 bis Bd Arago 75014 Paris, France
}

\author{Michel Cass\'e}
\address{
Service d'Astrophysique, DAPNIA, DSM, CEA, Orme des Merisiers,
91191 Gif sur Yvette CEDEX France and\\
Institut d'Astrophysique de Paris, 98 bis Bd Arago 75014 Paris, France
}

\author{Marc Rabiet}
\address{
Institut d'Astrophysique de Paris, 98 bis Bd Arago 75014 Paris, France
}

\maketitle

\begin{abstract}
We update Standard Big Bang Nucleosynthesis (SBBN) calculations on the basis of
recent nuclear physics compilations (NACRE in particular), experimental and
theoretical works.
By a Monte--Carlo technique, we calculate the uncertainties on the light
element yields (\qua, \deu, \tro\ and \sep) related to nuclear reactions.
The results are compared to observations that are thought to be
representative of the corresponding primordial abundances. 
It is found that \sep\ could lead to more stringent constraints on the
baryonic density of the universe (\obh) than deuterium, because of much 
higher observation statistics and an easier extrapolation to primordial values.
The confrontation of SBBN results with \sep\ observations 
is of special interest since other independent approaches
have also recently provided \obh\ values:
{\em i)} the anisotropies of the Cosmic 
Microwave Background by the BOOMERANG, CBI, DASI and MAXIMA experiments 
and {\em ii)} the \lyma\ forest at high redshift.
Comparison between these results obtained by different methods provides a
test of their consistency and could provide a better determination of the
baryonic density in the universe.
However, the agreement between \obh\ values deduced from SBBN calculation
and \sep\ observation on the one hand and CMB observations on the other hand
is only marginal.
\end{abstract}


\section{Introduction}

Recently, different ways to determine the baryonic density of
the universe have been exploited. 
Here, we use the usual notations where \ob\ denotes 
the ratio of the baryon density over the critical density of the universe,
and $\eta$ is the baryon over photon ratio. They are related by 
\obh=3.65$\times10^7\;\eta$ with $h$ the Hubble constant in units of 
100~km$\cdot$s$^{-1}$$\cdot$Mpc$^{-1}$.
It is now possible to confront the results of these different approaches to
test the validity of the underlying model hypothesis and hopefully
obtain a better evaluation of this crucial cosmological parameter.
Three independent methods have been used so far to derive the baryon
density of the universe, {\it i)} the pioneering one, (Standard)
Big-Bang Nucleosynthesis (SBBN), based on nuclear physics in the early
universe, {\it ii)} very recently, the study of the Cosmic Microwave
Radiation (CMB) anisotropies and {\it iii)} the census of H (and also He$^+$)
atomic lines from the \lyma\ forest at high redshift.

Regarding the uncertainties attached to each of these methods, those
related to SBBN are probably the best controlled. Standard BBN depends
essentially on one parameter, $\eta$, the baryon to photon ratio since 
the number of light neutrinos is essentially known.
This model, in its standard version, has survived for many decades
showing its robustness.
It success in reproducing the light element (\qua, \deu, \tro\ and \sep)
primordial abundances over a span of 10 orders of magnitude is 
remarkable\cite{Sch98}.
The leading uncertainties come from the observations of the isotopes
in different astrophysical sites and the way they are interpreted
(\deu\ in particular) to estimate the primordial
abundances and the insufficient knowledge of some reaction rates.
Lithium suffers from two drawbacks {\it i)} it is affected more
than any other light isotope by uncertainties in the nuclear reaction rates
{\it ii)} the valley shape in its abundance versus $\eta$ curve leads to
two possible $\eta$ values for a given abundance.
This shape is due to its production modes, by $^3H$+\qua\ and \tro+\qua,
respectively at low and high baryonic density.
The first difficulty could be reduced by a better determination of a few
key cross sections, but the second one is intrinsic to the calculation.
Thus to remove the
degeneracy on the baryon density, lithium should be associated to, at least,
one other light element, deuterium for instance.
However, the relation between \sep\ observations and its primordial
abundance seems more straightforward than for \deu.

In the case of the CMB, the \obh\ values deduced from observations tend
to converge but their interpretations are probably still model dependent.
The CMB analyses involve many parameters, principally the various energy
densities ($\Omega_{tot}$, \ob, $\Omega_\Lambda$, respectively the total 
density, baryonic density and the cosmological constant contribution), $h$, 
the initial fluctuation spectrum index ($n_s$), the reionisation optical
depth ($\tau_c$) and the overall normalization.
The baryon density is extracted from the amplitudes of the acoustic
peaks in the angular power spectrum of the CMB anisotropies.
It is important to note that the ratio of amplitudes between the first
and second peaks increases with \ob, in contrast with all other
cosmological parameters.
Hence, the determination of \obh\ does not suffer from the cosmic
degeneracy that affects $\Omega_\Lambda$ and 
$\Omega_m$ and a high precision can be expected\cite{Hu97}. 
However, these values are obtained in the framework of inflationary models 
that could be altered in other cosmological contexts\cite{Bou00}.

The third method is based on the study of the atomic HI and HeII \lyma\ 
absorption lines
observed in the line of sight of quasars. Quasars being the brightest objects
of the universe, they can be observed at very large redshift.
On their line of sight, atoms both in diffuse or condensed structures absorb
part of their radiation, making absorption lines apparent.
It allows in particular to study the intergalactic medium via the so-called
\lyma\ forest.
This method leads to an estimate of the baryon content of the Universe on large
scales\cite{Rie98,Wad00}.
Indeed, the evaluation of \obh\ through the study of the evolution of the 
\lyma\ forest in the redshift range 0$<z<$5, though indirect because of the
relatively large ionisation uncertainties, 
leads to results consistent with the two previous methods.

In the following, we update the Big Bang Nucleosynthesis calculations on
the basis of the recent NACRE compilation\cite{NACRE} of reaction rates
supplemented by other recent works\cite{Smi93,Bru99,Che99}.
We performed Monte--Carlo calculations to estimate the uncertainties on light
element yields arising from nuclear reactions alone.
Similar calculations have been performed recently, based on a different
compilation and analysis of nuclear data (Nollett and Burles\cite{NB00},
and  Cyburt et al.\cite{Cyb01}).
However, here, we put the emphasis on \sep\ as its primordial abundance
is more reliable than that of other light isotopes (\deu\ in particular).
Using \sep\ as the main {\em baryometer} one deduces \obh\ and we compare it
with {\it i)} with the helium and  deuterium primordial abundances,
{\it ii)} other SBBN investigations and {\it iii)} independent evaluations
(CMB and \lyma\ forest).

\section{Observational constraints from the light elements}
\label{s:obs}

Here we present the selection of astrophysical observations that we use 
for the determination of the baryon density of the Universe from the SBBN
calculation.
To estimate primordial abundances, observations are made on the oldest
objects that are characterized by their high redshift $z$ or low {\em
metallicity}
\footnote{
Metallicity represents the abundance of {\em metals} which, in the
astrophysical language, corresponds to all elements above helium. As
metallicity increases in the course of galactic evolution, this is an
indicator of the age of an object.
Abundances of common elements like Fe (or e.g. Si), are often taken as
representative of the metallicity.
The notation
$[Fe/H]\equiv\log\left(Fe/H\right)_{\mathrm star}-\log\left(Fe/H\right)_\odot$
is often used.
For instance, $[Fe/H]$ = -2 corresponds to 1\% of the solar ($\odot$)
metallicity.
Otherwise, $D/H$ or $^7Li/H$ for instance, represent the ratio of abundances
by number of atoms.
}.

The determination of the primordial \qua\ abundances is derived from 
observations of metal--poor, extragalactic, ionized hydrogen (HII) regions.
This extraction is difficult to the level of precision required, due to
the incomplete knowledge of the different atomic parameters involved.
Olive and Skillman\cite{Oli01} have studied in great detail the systematic
uncertainties, and concluded that the typical errors given in previous
studies are underestimated by a factor of about two.
The extreme values published\cite{Fie98,Izo98,Pei00} cover the range
fom 0.231 to 0.246 (in mass fraction), putting little constraint on models.
Consequently, in this work, \qua\ will not be considered as a discriminating
indicator of the baryonic density.

\begin{figure}
\begin{center}
\epsfig{file=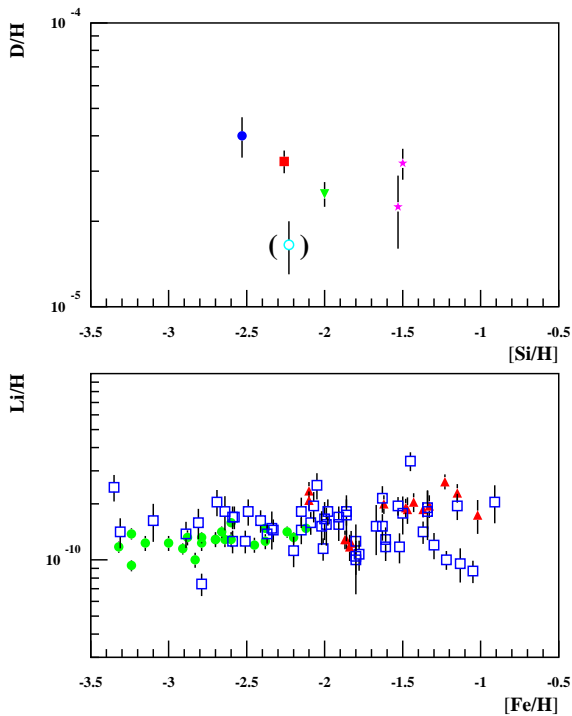,width=8.cm}
\caption{Observed abundances as a function of metallicity from objects 
which are expected to reflect primordial abundances.
Upper panel : observed \deu\ abundances, from
Refs.~\protect\cite{Bur98,OMe01,Pet01,DOd01,Lev01}
(stars corresponding to two analyses of observations of the same 
cloud\protect\cite{DOd01,Lev01}).
Lower panel : observed \sep\ abundances, circles\protect\cite{Rya01} and
triangles\protect\cite{Rya99} from Ryan et al.; squares from Bonifacio and   
Molaro\protect\cite{Bon97}.}
\label{f:lid}
\end{center}
\end{figure}

Deuterium is peculiar because, after BBN, this fragile isotope, can, in
principle, only be destroyed in subsequent stellar or galactic nuclear
processing.
Hence the primordial abundance should be represented by the {\em highest}
observed value.
It is measured essentially in three astrophysical sites, {\it i)}
in the local interstellar medium (present value), {\it ii)} in the
protosolar cloud (4.6 Gyr ago) and
{\it iii)} in remote cosmological clouds on the line of sight of high
redshift quasars (large lookback time).
In principle, the later sample {(\it iii)} should be the closest
representative of the primordial \deu\ value, but up to now, the
observations lead to two ranges of \deu\ abundance values.
However, very recently, it has been shown\cite{Kir01} that the \deu\
abundance on the
line of sight of the QSO~PG1718+4607 quasar cannot be  determined due to
blending between the \lyma\ and the main hydrogen absorption lines,
contrary to a previous study\cite{Web97}.
Since this observation was the main evidence for a high $D/H$ value
($\sim10^{-4}$ corresponding to a low $\eta$ range; see for instance
Ref.~\cite{Van00a}), the very high primordial \deu\ abundance seems to have
lost its support and hence only one range of (low) $D/H$ values remains.
However, the \deu\ abundance data from cosmological clouds remain scarce and
scattered (Fig.~\ref{f:lid}, upper panel).
The extreme values deduced from the different
observations\cite{Bur98,OMe01,DOd01,Pet01} lead to the interval
$1.3\times10^{-5}<D/H<4.65\times10^{-5}$ (including error bars).
This dispersion (amounting to a factor of about 3), if physical and not
observational, casts a doubt on the direct identification of the
observed values with the primordial \deu\ abundance.
Alternatively, it could indicate that this fragile isotope has already
been processed in these high redshift clouds despite their low 
metallicity\cite{Fie01}.
Thus, in this perspective, averaging the $D/H$ abundances measured in
cosmological clouds to infer the primordial value seems somewhat
inappropriate.
In addition the observations of the absorbing cloud on the line of sight of 
QSO~0347-3818 have been analyzed using two methods\cite{DOd01,Lev01}
and lead to different values (stars in Fig.~\ref{f:lid}).  
This suggest that systematic errors may still be important.

%  Position of Figure 1

On the other hand, it has been shown\cite{Vid01} that there
exists a large dispersion in the local measurements ($0.5$ to
$4.\times10^{-5}$).
This could indicate that unknown processes are at work to modify the \deu\
abundance at small scale in our Galaxy.
Thus, if it is confirmed that local \deu\ abundances are scattered as the
result of yet unknown physical processes, the same thing could occur
in absorbing clouds at large redshift.
In addition, the lowest value obtained at high redshift\cite{Pet01}
($D/H$ = $1.65\pm0.35\times10^{-5}$)
is uncomfortably close to both the solar system
and interstellar ones (respectively around $2.1\times10^{-5}$ and
$1.5\times10^{-5}$\cite{Gei98,Lin98,Vid01}).
It is inconsistent with even the most conservative galactic evolution models
since it would require a negligible \deu\ destruction.
Note that this lowest value\cite{Pet01} (in parenthesis in
Fig.~\ref{f:lid}) is affected by a large uncertainty concerning the
level of the continuum and the blending of the relevant
lines\cite{Petitjean}.
If this questionable observation is put aside, a trend appears in the
$D/H$ data versus metallicity $[Si/H]$ (Fig.~\ref{f:lid}) showing
a \deu\ abundance decreasing when metallicity increases,
as qualitatively expected from stellar evolution\cite{Fie01}.
(To view the trend we do not consider the alternative analysis\cite{Lev01} 
of QSO~0347-3818 to be consistent with the analyses of the other observations.)
Accordingly, the true primordial $D/H$ value should be obtained from
extrapolation to zero metallicity.
Consequently, even though, in principle \deu\ better constrains $\eta$ 
than \sep\ (U--shape of the curve), the value of its primordial abundance
is still a matter of debate.
We adopted the highest observed value\cite{Bur98} in a cosmological cloud
assuming that lower values are the result of
subsequent processing.

Compared to \deu, the determination of the \sep\ primordial abundance from
observations leaves less room for interpretations.
Since the discovery of a plateau in the lithium abundance as a function of
metallicity (Fig.~\ref{f:lid}, lower panel), drawn for metal poor dwarf 
stars\cite{Spi82}, many new observations have strengthened its existence.
The fact that the abundance does not increase with time (metallicity) at the
surface of the oldest stars was interpreted as being representative of
the primordial \sep\ abundance\cite{Bon97,Rya99,Rya01}.
As such, these measurements could have been affected by two processes
{\em i)} a production related to non thermal (spallation) nuclear reactions
(mainly $\alpha+\alpha$) in the interstellar medium, increasing the
amount of lithium in forming stars at a given metallicity and {\em ii)} a
depletion in the envelope of these stars.
The contribution of the
first process is small at very low metallicity amounting
typically to less than 10\% at a metallicity $[Fe/H]$=-2\cite{Van99}.
At $[Fe/H]$=-1, this contribution is more significant but remains within the
dispersion of the data.
The second one concerns the potential depletion of lithium by nuclear
destruction and possibly by diffusion and rotational mixing\cite{The01}.
The small
scatter of the data, over three metallicity decades, on the one hand and
the presence of the even more fragile \six\ isotope in a few halo stars
(e.g. Ref.~\cite{Cay99}) on the other hand, strongly limit the amount of
possible depletion. Hence, this effect should also be within the dispersion
of the data.
Bonifacio  and Molaro\cite{Bon97} have deduced from their large observational
sample a primordial value : $\log(Li/H)$ = -9.762$\pm$0.012 
(statistic, 1$\sigma$) $\pm$0.05 (systematic).
More recently, Ryan et al.\cite{Rya99,Rya00}, on the basis of their 
observations, have provided a new determination:
$^7Li/H$ = $(1.23^{+0.68}_{-0.32})\times10^{-10}$.
Their mean value and (95\% confidence) limits take into account all possible
contributions from \sep\ depletion mechanisms and bias in analysis.
But the main difference with the earlier work\cite{Bon97} is that they 
have taken into account a slight rise of the lithium abundance due to
spallation reaction leading to a smaller primordial abundance when
extrapolated a zero metallicity.
Accordingly, we adopt their range for the primordial \sep\ abundance.

\section{CMB and Lyman--$\alpha$ forest observations}
\label{s:cmb}

CMB anisotropy measurements give independent estimates of the baryonic
density of the Universe.
The first determinations  of \obh\ from BOOMERANG and MAXIMA
\cite{Bal00,Lan01,Jaf01} yielded \obh$\approx$0.03.
The  Cosmic Background Imager (CBI), ground based, has given preliminary
results in marked contrast\cite{Pad01} (\obh$\approx$0.009) to
BOOMERANG/MAXIMA values.
The Degree Angular Scale Interferometer, DASI (along with its sister
instrument CBI)\cite{Pry01}, is one of the new compact interferometers
specifically built to observe the CMB. Combined with the large angle
measurements made by COBE, it has been able to reveal a
significant signal in the second peak region and has determined
\obh=0.022$^{+0.004}_{-0.003}$ (1$\sigma$).
Recently, new analyses\cite{Net01,Ber01} of BOOMERANG data have also led to
\obh\ = 0.022$^{+0.004}_{-0.003}$ (1$\sigma$).
But the situation is not yet settled and a  wealth of new data is
expected from future ground instruments, long balloon flights and especially 
satellites (MAP, PLANCK--SURVEYOR).

As mentioned above, the study of the baryon content of the intergalactic
medium evolution of the \lyma\ forest in the redshift range
$0<z<5$ leads to an evaluation of \obh. Such analyses have lead to
\obh$\ge$0.0125\cite{Rie98,Petitjean} and $\le$0.03 
\cite{Rie98,Wad00,Petitjean}. 
Hui et al.\cite{Hui01} using recent observations\cite{Sco00,Ste01}
have found \obh=0.03$\pm$0.01.
These various results are summarized in Table~\ref{t:ob}, compared in
Fig.~\ref{f:lik} and discussed in Sect.\ref{s:res}.

\section{Nuclear data}
\label{s:nucl}

Most of the important reactions for \sep\ production (Table~\ref{t:reac})
are available in the NACRE compilation of thermonuclear reaction
rates\cite{NACRE}.
Other reactions in our BBN network are adapted from an earlier
compilation\cite{Smi93} or more recent works\cite{Bru99,Che99}.
In our previous studies, we have studied the influence of individual
reactions\cite{Van00a} or extreme yield limits\cite{Van00n} obtained
when considering all combinations of low and high rates.
Here, instead, we have performed Monte-Carlo calculations to obtain 
statistically better defined limits, as we did in a previous work\cite{Coc00}
that was limited to the NACRE reactions.  
Here we update these calculations by taking into account uncertainties on
the remaining reactions. In the following, we discuss the origin of these
calculated uncertainty limits.

One of the main innovative features of NACRE with respect to former
compilations\cite{CF88} is that  uncertainties are analyzed in detail and
realistic lower and upper bounds for the rates are provided.
Using these low and high rate limits, it is thus possible to
calculate the effect of nuclear uncertainties on the light element yields.
Recently, two other SBBN calculations have been performed: one based on the 
Nollett and Burles\cite{NB00} compilation (hereafter NB) and another 
on a partial reanalysis the NACRE data by Cyburt et al.\cite{Cyb01}
(hereafter CFO).
The NB and NACRE compilations differ in several aspects.
The NB compilation addresses primordial nucleosynthesis while NACRE
is a general purpose compilation. Consequently, NB contains a few
more reactions of interest to BBN and a few more data in the energy range of
interest. Also, {\em from the statistical point of view}, the rate
uncertainties are better defined in NB, however the astrophysical
$S$--factors\footnote{The (astrophysical $S$--factor is defined by
$\sigma(E)\equiv {{S(E)}\over{E}}
\exp\left(-2\pi\eta\right)\equiv {{S(E)}\over{E}}
\exp\left(-\sqrt{{E_G}\over{E}}\right)$ where here $\eta$
$\left(=Z_1Z_2e^2/{\hbar}v\right)$ is the Sommerfeld
parameter. It reduces the strong dependency of the cross
section ($\sigma$) at low energy by approximatively correcting for the
penetrability of the coulomb barrier.}
are fitted by splines which have no physical justification and can
produce local artifacts by following to closely experimental data points.
On the contrary, the NACRE compilation spans wider energy ranges, and over
these ranges, the $S$--factors are fitted to functions based on theoretical
assumptions.

%  Position of Figure 2
\begin{figure}
\begin{center}
\epsfig{file=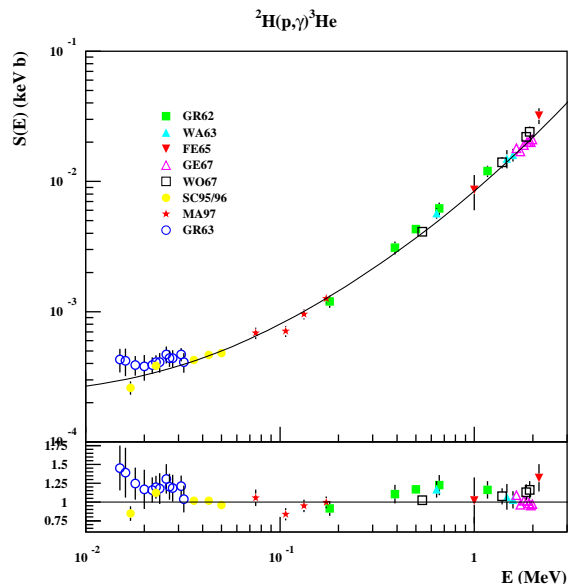,width=8.cm}
\caption{Upper panel: astrophysical $S$--factor for the $^2$H(p,$\gamma)^3$He
reaction (adapted from NACRE\protect\cite{NACRE}, for references see NACRE
except for SC96\protect\cite{Sch96}.)
Lower panel: relative dispersion of residuals 
($S_{\mathrm exp}/S_{\mathrm fit}$).}
\label{f:dpg}
\end{center}
\end{figure}

\begin{figure}
\begin{center}
\epsfig{file=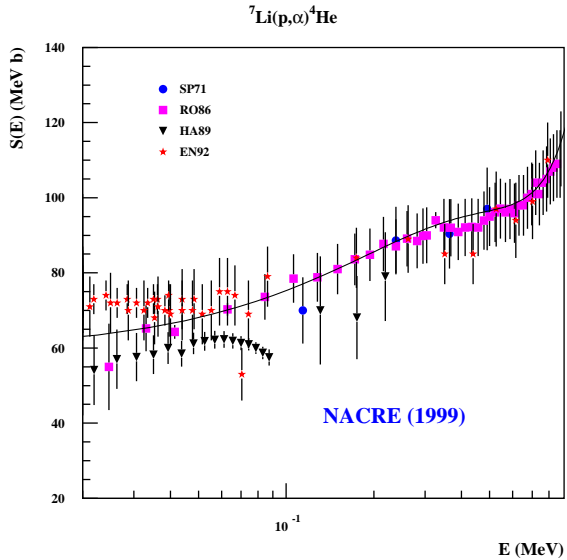,width=8.cm}
\caption{Astrophysical $S$--factor for the $^7$Li(p,$\alpha)^4$He reaction
(adapted from NACRE\protect\cite{NACRE}). For references other than 
HA89\protect\cite{HA89} and RO86\protect\cite{RO86}, see NACRE.}
\label{f:li7pa}
\end{center}
\end{figure}

Fig.~\ref{f:dpg} shows that outside of resonances a simple fit (second order
polynomial in this case) is sufficient to account for $\approx$2 orders of
magnitude variation in $S$.
For the $D(p,\gamma)^3He$ reaction, as in Ref.~\cite{Van00a}, we use the
revised data of Schmidt et al.\cite{Sch96} (SC96 in Fig.~\ref{f:dpg})
not considered in NACRE.
The lower panel displays the ratio between the
experimental and fitted $S$ values. The dispersion around unity is small and
can hardly be considered physical. It should be noted that by using such a
simple fit (when permitted by theory), precise data points outside the
range of BBN energy (e.g. the high energy data point in this case) helps
constrain the fitted $S$--factor in the region of interest ($\sim$0.1~MeV in
this case).
This eliminates spurious local effects induced by a few erratic
data points associated with experimental problems rather than a genuine
physical effect.
For instance, Fig.~\ref{f:li7pa} (adapted from NACRE\cite{NACRE}) shows
the nuclear data and the NACRE recommended $S$--factor for the
$^7$Li(p,$\alpha)^4$He reaction can be compared with Fig.~12 of
NB\cite{NB00}. In this latter analysis, the $S$--factor is unduly influenced
by the Harmon\cite{HA89} data which is a measurement relative to the
assumed constant $S$--factor of the $^6$Li(p,$\alpha)^3$He reaction.
The rise of $S$ at low energy is also more likely interpreted by the effect
of atomic electron screening of the nucleus. In this energy range, the cross
section is expected to be non resonant, mainly determined by the tails of
higher energy resonances. This is why, NACRE has constrained $S$ to follow
a low order polynomial with the consequence
that the good and extensive data provided by Rolfs and Kavanagh\cite{RO86}
are better taken into account.

\begin{table}
\caption{Influential reactions and their sensitivity to nuclear uncertainties
for the production of \qua, \deu, \tro\ and \sep\ in SBBN.}
%\tiny
\begin{flushleft}
\begin{tabular}{|l|c|c|c|c|}
\hline
Reaction~${\backslash}$~${\Delta}N/N$
& $^4$He&D&$^3$He&$^7$Li\\
\hline
$^1$H(n,$\gamma)^2$H$^{(a}$&n.s.&n.s.&n.s.&0.08\\
\hline
$^2$H(p,$\gamma)^3$He&n.s.&-0.19&0.19&0.26\\
\hline
$^2$H(d,n)$^3$He&n.s.&-0.09&0.06&0.12\\
\hline
$^2$H(d,p)$^3$H&n.s.&-0.03&-0.04&0.01\\
\hline
$^3$H(d,n)$^4$He&n.s.&n.s.&n.s.&-0.07\\
\hline
$^3$H($\alpha,\gamma)^7$Li&n.s.&n.s.&n.s.&0.24\\
\hline
$^3$He(n,p)$^3$H$^{(b,d}$&n.s.&n.s.&-0.06&-0.03\\
\hline
$^3$He(d,p)$^4$He$^{(c,d}$&n.s.&n.s.&-0.12&-0.12\\
\hline
$^3$He($\alpha,\gamma)^7$Be&n.s.&n.s.&n.s.&0.39\\
\hline
$^7$Li(p,$\alpha)^4$He&n.s.&n.s.&n.s.&-0.25\\
\hline
$^7$Be(n,p)$^7$Li$^{(c,d}$&n.s.&n.s.&n.s.&-0.13\\
\hline
\end{tabular}
%\normalsize
${\Delta}N/N{\equiv}N_h/N_l-1$\\
n.s. : not significant ($|{\Delta}N/N|<0.01$).\\
$^{a)}$ Chen \& Savage \protect\cite{Che99};
$^{b)}$ Brune et al. \protect\cite{Bru99};
$^{c)}$ Smith, Kawano \& Malaney\protect\cite{Smi93}.\\
$^{d)}$ $\pm$1$\sigma$ variation.
\end{flushleft}
\label{t:reac}
\end{table}

In some cases, [e.g. the D$(d,n)^3$He reaction] the NB compilation provides 
more data points in the region of interest than NACRE.
However, considering a wider energy range, NACRE relies on an
interpolation between high and low energy data.
It is difficult to further compare the reaction rates obtained in both 
compilations because NACRE provides reaction rate limits (and 
in some cases $S$--factors) that can be used for e.g. subsequent
Monte--Carlo calculations, while in NB the rate
calculation and Monte-Carlo cannot be disentangled.    
Indeed, in NB, the Monte-Carlo procedure is not applied to the rates but
to the data points within experimental errors followed by spline fitting. 
This method is expected to take better into account experimental errors 
but is difficult to evaluate especially because it could depend on
the partitions of the energy interval used for spline fitting which are
not given.
Nevertheless the final results (i.e. \qua, \deu, \tro\ and
\sep\ yields) are in good agreement\cite{Coc00} showing that both approaches
are valid.
CFO\cite{Cyb01} have reanalyzed the compiled data from NACRE using
the same $S$--factor energy dependences but leaving the scaling factors free.
They found that global normalization factors were slightly different from
the NACRE ones. As the fitting procedure is straightforward, the origin for
this difference is difficult to interpret. One possibility is that a few data
points were excluded by NACRE due to suspected experimental problem or
physical bias (screening at low energy). CFO\cite{Cyb01} subsequently 
determined scaling factors for each experiment to take into account 
systematics. 
From the dispersion of these factors they obtained a better evaluation of
rate uncertainties including systematic effects.
These analyses underestimate the fact that data on experimental cross
sections are in general of much better quality at BBN energy or above than
at lower energies.
Indeed the cross sections for charged particle reactions drop very
rapidly at low energy (Coulomb barrier) making experiments more and more
difficult and hence subject to systematic errors. In addition screening of
the nucleus by atomic electrons is known to affect cross sections at low
energy\cite{Ang93}.
Hence, a scaling factor obtained from a low energy
measurement is likely to be more affected by systematic errors that another
one derived from a high energy data set, even if the quality of the fits is
the same. This could affect the calculated CFO systematic error contribution.

%  Position of Figure 3

Clearly those recent compilations or analyses\cite{Smi93,NACRE,NB00,Cyb01}
have all improved the determination of BBN rates and rate uncertainties but
more progress could be made. This would include a better theoretical
determination of the energy dependence of $S$--factors sometimes just
assumed to be a low order polynomial. This would improve the constraint given
by the shape of $S(E)$. Such work is under way\cite{Des01}.

Due to the difficulty of defining a universal
statistical method for the wide set of reactions (each with its
peculiarities) and range of temperature, the NACRE rate limits correspond
to upper and lower bounds rather than standard deviations.
Accordingly, we assumed a uniform distribution for the rates between the
limits (keeping the mean rates equal to the recommended rates).

Four reactions of interest to BBN are not found in the NACRE compilation of
{\em charged} particle induced reactions. They are the neutron induced
reactions n$\leftrightarrow$p, $^1$H(n,$\gamma)^2$H, $^3$He(n,p)$^3$H,
$^7$Be(n,p,)$^7$Li and also $^3$He(d,p)$^4$He.
The first reaction governs the neutron--proton ratio at the time of
freeze--out and hence directly the \qua\ primordial abundance.
The main source of uncertainty on this rate used to be the neutron 
lifetime but its value is now precisely (886.7$\pm$1.9~s) known\cite{PDG}.
So even though improved theoretical calculations\cite{Bro01} may introduce
small corrections, this reaction is now sufficiently known for BBN
calculations.
The following reaction, $^1$H(n,$\gamma)^2$H, also relies almost
exclusively on theory. A new calculation including quoted uncertainties
has been made available recently\cite{Che99}. For this reaction, the
uncertainties arising from the experimental input data (one low--energy
normalization value for the cross section) are expected to be
much smaller than those from theory. The errors are given by the order
of the first neglected terms in the expansion\cite{Che99}. To derive the
reaction rate and its limits, we performed numerical integrations using the
analytical formulas for the cross section and its calculated
uncertainties\cite{Che99}. As there is no way to determine the statistical
distribution of these {\em theorical errors} we adopted (as for NACRE) a
uniform distribution as for the following reaction.
Following discrepancies and lack of documentation for the $^3$H(p,n)$^3$He
reaction and its inverse, a new and precise measurement has recently been
performed\cite{Bru99}. The results corroborate the cross section provided
by the ENDF/B-VI evaluation of neutron data\cite{ENDF} leading to an
estimated uncertainty of 5\%\cite{Bru99}.
For the two remaining reactions $^3$He(d,p)$^4$He and $^7$Be(n,p)$^7$Li
no new measurements are available and we adopt accordingly the reaction rate
and uncertainties provided by Smith, Kawano and Malaney\cite{Smi93}.
(In the more recent NB analysis the reaction rates and uncertainties are not
available due to the intricate coupling of the fitting and Monte--Carlo
methods.)
They performed an R-matrix analysis (a standard nuclear physics method to
tackle resonant reactions) complemented by polynomial fits to the data.
Their quoted 1$\sigma$ uncertainties are respectively 8 and
9\%\cite{Smi93} and we use accordingly for these two reactions a gaussian
distribution of errors.

%  Position of Table 1

%  Position of Figure 4
\begin{figure}
\begin{center}
\epsfig{file=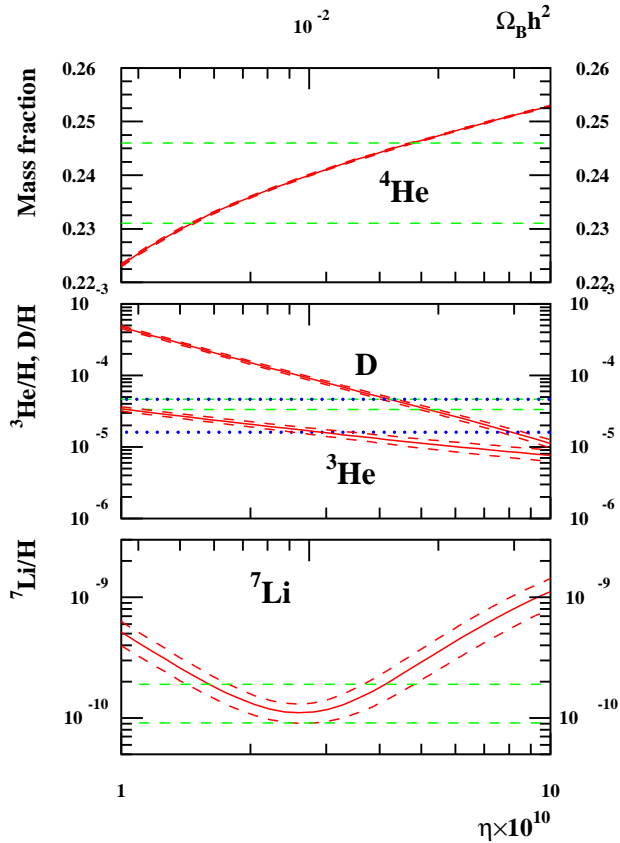,width=8.cm}
\caption{Abundances of \qua\ (mass fraction), \deu, \tro\ and \sep\ (by
number relative to H) as a function of the baryon over photon ratio 
$\eta$. Mean values (solid curves) and 2$\sigma$ limits (dashed curves) 
are obtained from Monte Carlo calculations. 
Horizontal lines represent primordial \qua, \deu\ and \sep\ abundances deduced
from observations (see text).
For \deu\ the dotted lines represents the range of observed values (see
Fig.~\protect\ref{f:lid}) while the dashed lines corresponds to the adopted 
value of Ref.\protect\cite{Bur98}. 
}
\label{f:bbn}
\end{center}
\end{figure}

\section{Results}
\label{s:res}

In a first step, we complemented our previous analysis\cite{Van00a}, limited
to NACRE reactions, by calculating the influence of individual reaction rates
on \qua, \deu, \tro\ and \sep\ yields.
Then we calculated the maximum of the quantity
${\Delta}N/N{\equiv}N_{high}/N_{low}-1$ within the range of $\eta_{10}$
variations for each of the 4 isotopes. Positive (resp. negative) values
correspond to higher (resp. lower) isotope production when the high rate
limit is used instead of the low one (see Ref.~\cite{Van00a}).
Results are displayed in Table~\ref{t:reac}.

\begin{figure}
\begin{center}
\epsfig{file=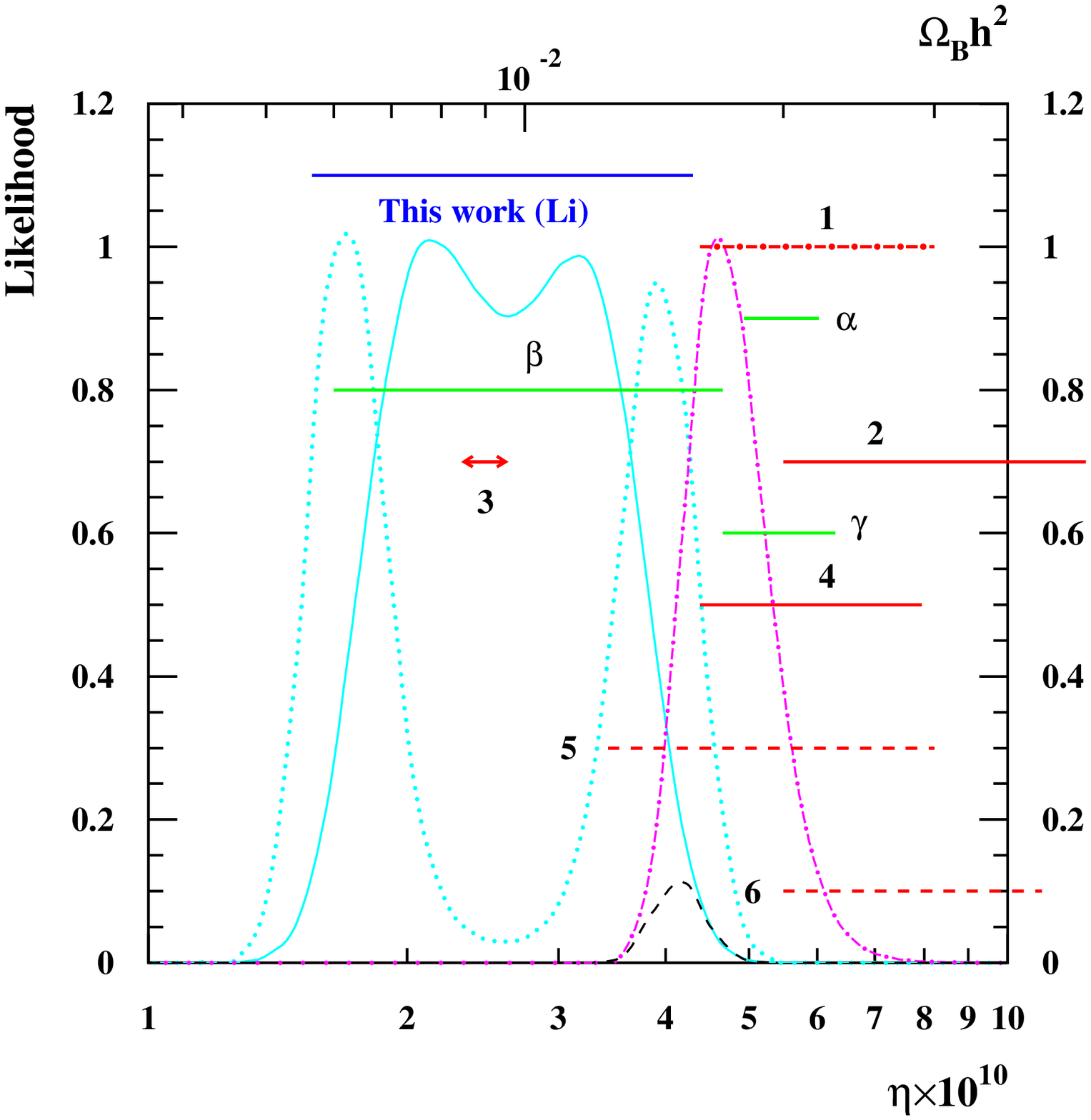,width=8.cm}
\caption{Likelihood function for \sep, \deu,
\sep+\deu\ and $\eta$ ranges from CMB, \lyma\ and other SBBN calculations.
The horizontal lines represent \obh\ and $\eta$ intervals :
2$\sigma$ confidence limits (solid lines), twice the 1$\sigma$ limits
(dash--dotted line) or statistically
less well defined limits (dashed lines and arrows).
Labels ($\alpha$, $\beta$, $\ldots$, 5, 6) point to the last
column of Table~\protect\ref{t:ob} were more details are given.  
The solid curve represents ${\mathcal L}^7$ for the adopted value of \sep\
primordial abundance while the dotted curve display, 
${\mathcal L}^7$ if the higher value of Bonifacio and
Molaro\protect\cite{Bon97} is adopted.
The likelihood function for \deu\ (${\mathcal L}^2$, dash dotted curve)
is only marginally consistent with ${\mathcal L}^7$ as shown by
${\mathcal L}^{7,2}$ (dashed curve). [${\mathcal L}^2$ and ${\mathcal L}^7$
have been normalized to ${\mathcal L}_{\mathrm max}=1$.]
}
\label{f:lik}
\end{center}
\end{figure}

Then we performed Monte--Carlo calculations with the rate distributions
discussed above.
For each $\eta$ value, we calculated the mean value and standard deviation
($\sigma$) of the \sep\ yield distribution. The corresponding
$\pm$2$\sigma$ limits are represented in Fig.~\ref{f:bbn}.
In 
Fig.~\ref{f:lik} is represented the likelihood functions for \sep\ only 
[${\mathcal L}^7(\eta)$], \deu\ only [${\mathcal L}^2(\eta)$] and for both
[${\mathcal L}^{7,2}(\eta)$].  
The $n\sigma$ confidence intervals are obtained by solving:
$\ln\left(\mathcal{L}(\eta)\right)=
\ln\left(\mathcal{L}_{\mathrm max}\right)-n^2/2$, for $\eta$.
To calculate ${\mathcal L}$, we use the abundance distributions
obtained by Monte--Carlo together with a normal distribution associated 
with the adopted primordial abundances.
Following the conclusions of Sect.~\ref{s:obs}, we assumed that the
primordial \sep\ abundance is such that $\log(^7Li/H)$ is normally
distributed with mean -9.91 and standard deviation $\sigma$ = 0.19/2
as given by Ryan et al.\cite{Rya00}.
We neglected the asymmetry in the error bars
(-9.91$^{+0.19}_{-0.13}$\cite{Rya00}) by taking the largest because the
smallest concerns the \sep\ lower limit that only affects the
${\mathcal L}^7$ central dip, and not the $\eta$ limits (Fig.~\ref{f:bbn}
lower panel).
Unfortunately, the
U--shape of the \sep\ curve together with the Ryan et al.\cite{Rya00} values
leads to a merging of the low and high $\eta$ intervals. For comparison, we
also show the likelihood function obtained when using the Bonifacio and
Molaro\cite{Bon97} older value exhibiting two $\eta$ intervals. Their
merging clearly originates from the new lower primordial abundance obtained
when correcting for the apparent $^7Li/H$ versus metallicity slope (see
Fig.~\ref{f:lid} and Sect.~\ref{s:obs}).
From this curve, we obtain for \obh, the range 0.006--0.016 (95\% c.l.,
\sep\ only).

To calculate ${\mathcal L}^2$, the likelihood function concerning \deu\ only,
we adopted the highest observed value\cite{Bur98} (see Sect.~\ref{s:obs}).
The ${\mathcal L}^2$ curve (Fig.~\ref{f:lik}) is centered on
$\eta$=5$\times10^{-10}$ and is only marginally consistent with the
${\mathcal L}^7$ one. The situation is even worse if the smaller values of the
\deu\ primordial abundances are considered. 
The agreement between SBBN and
the \sep\ and \deu\ observed primordial abundances is impressive when
considering the orders of magnitude involved but remains only moderately
good when trying to determine the \obh\ value.
The global ${\mathcal L}^{2,7}$ likelihood function provide
\obh=0.015$\pm$0.003 (2$\sigma$) which is reasonably compatible with the
BOOMERANG\cite{Ber01} (\obh=0.022$^{+0.004}_{-0.003}$; 1$\sigma$),
and DASI\cite{Pry01} (\obh=0.022$^{+0.007}_{-0.006}$; 2$\sigma$) values.
However, as discussed above this is based on unsettled discrepancies on the
ranges dictated separately by \sep\ and \deu.
Better agreement between SBBN and CMB has been claimed in other 
works\cite{Bur01}:
\obh=0.020$\pm$0.002 (95\% c.l.). This result from the choice of smaller
\deu\ primordial abundances, drives \obh\ to higher values but at the
expense of compatibility with \sep.
Considering that the chemical evolution of \deu\ from BBN to present is not
well known, but given that it can only be destroyed in this process, our choice of
adopting the higher observed value seems justified.
However, a better compatibility with the more reliably determined primordial
\sep\ abundance, would even favor a possible higher \deu\ abundance, implying
that \deu\ has already been partially destroyed in the cosmological cloud. 
Of course this would decrease the compatibility with CMB observations.

%  Position of Figure 5

%  Position of Table 2

\begin{table}
\caption{Comparison of \obh\ from different methods. Limits are given for
2$\sigma$ except for values in italic (see table footnotes).
The last column provides the labels for the \obh\ ranges displayed in
Fig.~\protect\ref{f:lik}.}
%\tiny
\begin{flushleft}
\begin{tabular}{|l|c|l|}
Method & \obh\ & Fig.~\ref{f:lik} \\
\hline
SBBN+Obs (Li), this work  & 0.006 -- 0.016 &\\ %2\sigma
SBBN+Obs (Li+D), this work  & 0.015$\pm$0.003 &\\ %2\sigma
SBBN  from Burles et al.$^{a)}$ & 0.020$\pm$0.002 & $\alpha$\\ % 95% cl
SBBN from CFO$^{b)}$ He + Li	& 0.006  -- 0.017 &$\beta$\\  % 95% cl
SBBN from CFO$^{b)}$ low D  &	0.017 -- 0.023 &$\gamma$    \\ % 95% cl
CMB BOOMERANG$^{c)}$	&
{\it 0.022}$^{\mathit +0.008}_{\mathit -0.006}$ &1 \\ % 1\sigma\times2
CMB from MAXIMA$^{d)}$ 	&	0.0325$\pm$0.0125 &2\\% 95% cl
CMB from CBI$^{e)}$	&      $\approx${\it 0.009} &3\\ % approximate
CMB from DASI$^{f)}$ &	0.022$^{+0.007}_{-0.006}$ &4\\%2\sigma
\lyma\ $^{g)}$ &  {\it 0.0125 -- 0.03} &5\\ % approximate
\lyma\ $^{h)}$	& {\it 0.03}${\mathit\pm0.01}$ &6\\ % approximate
\end{tabular}
$^{a)}$ Burles, Nollett \& Turner (2001) \protect\cite{Bur01}.\\
$^{a)}$ Cyburt et al. (2001) \protect\cite{Cyb01}.\\
$^{c)}$ de Bernardis et al. (2001) \protect\cite{Ber01};
2$\sigma$ interval approximated by twice the width of the 1$\sigma$ interval.
\\ % Boomerang
$^{d)}$ Stompor et al. (2001) \protect\cite{Sto01}.\\ % Maxima
$^{e)}$ Padin et al. (2001) \protect\cite{Pad01}; no confidence interval 
given.\\
$^{f)}$ Pryke et al. (2001) \protect\cite{Pry01}.\\ % DASI
$^{g)}$ Riediger et al. (1998) \protect\cite{Rie98}; estimated range of
values from Petitjean (2001) \protect\cite{Petitjean}.\\
$^{h)}$ Scott et al. (2000) \protect\cite{Sco00} and 
Hui et al. (2001) \protect\cite{Hui01}; no details on statistical
significance given.\\
\end{flushleft}
\label{t:ob}
\end{table}

\section{Conclusion}

Big Bang nucleosynthesis has been the subject of permanent interest since it
gives access to the baryon density which is a key cosmological parameter.
Though independent methods are now available, the SBBN one remains the most
reliable because {\it i)} the underlying physics is well known and {\it ii)}
there is essentially only one free parameter contrary to other methods.
It is worth pursuing the improvement of nuclear reaction rates and abundance
determination of light elements, essentially \deu\ and \sep.

Our SBBN results, \obh=0.006--0.016 based on \sep\ only and 0.015$\pm$0.003
with \deu\ are good agreement with those from Cyburt et al.\cite{Cyb01a}.
The \obh\ value derived by Burles, Nollett and Turner\cite{Bur01} (\deu\ and
\sep) is in reasonable agreement with ours. 
These results are broadly consistent with the CMB ones (MAXIMA, BOOMERANG 
and DASI) and those obtained via the observation of the \lyma\ forest at high
redshift (see also Cyburt et al.\cite{Cyb01a}).
However, the SBBN values derived separately from \sep\ and \deu\ are only 
marginally compatible and when using the more reliable indicator, lithium, the
agreement with the CMB and \lyma\ values is also marginal. 

It is interesting to note that the SBBN\cite{NB00,Van00a,Cyb01} results 
derived from the two recent and {\em independent} reaction rate compilations
(NACRE\cite{NACRE} and NB\cite{NB00}) agree very well.
Progress in the derivation of primordial abundances (\deu) are certainly 
needed, but improvement in the determination of nuclear reaction rates would
also be of interest (\sep\ nucleosynthesis specifically).
Concerning this last point, reanalysis of existing data constrained with
improved theoretical input are under way\cite{Des01}. However, it would be
even more important that new experiments dedicated to precise and systematic 
measurements (e.g. in Refs.\cite{RO86,Bru99}) of the lesser known 
cross section (Table~\ref{t:reac}) be undertaken.

\section{Acknowledgments}
We are deeply indebted to Roger Cayrel and Patrick Petitjean for
illuminating discussions.
We warmly thank Brian Fields and Keith Olive for the permanent
collaboration on the light elements.
We also thank Carmen Angulo and Pierre Descouvemont for clarifying
discussion on the NACRE methodology and Dave Lunney for a careful
reading of the manuscript.
This work has been supported by the PICS number 1076 of INSU/CNRS.

\end{document}